\documentclass[aps,a4paper,10pt,showpacs,prb,twocolumn]{revtex4-1}

\usepackage{amsfonts}
\usepackage{amsmath}
\usepackage{amssymb}
\usepackage{graphicx}
\usepackage{verbatim}
\usepackage{color}

\begin{document}

\title{Plasmon-mediated Coulomb drag between graphene waveguides}

\author{Artsem A. Shylau}
\email{arts@nanotech.dtu.dk}
\author{Antti-Pekka Jauho}

\affiliation{Center for Nanostructured Graphene (CNG), Department of Micro- and Nanotechnology, DTU Nanotech, Technical University of Denmark,
DK-2800 Kongens Lyngby, Denmark}

\date{\today }

\begin{abstract}
We analyze theoretically charge transport in Coulomb coupled graphene waveguides (GWGs).  The GWGs are defined using antidot lattices, and the lateral geometry bypasses many technological challenges of earlier designs.  The drag resistivity $\rho_D$, which is a measure of  the many-particle interactions between the GWGs,  is computed for a range of temperatures and waveguide separations.
It is demonstrated that for $T>0.1T_F$ the drag is significantly enhanced due to plasmons, and that in the low-temperature regime a complicated behavior may occur. In the weak coupling regime the dependence of drag on the interwaveguide separation $d$ follows $\rho_D \sim d^{-n}$, where $n \simeq 6$.
\end{abstract}


\maketitle

\section{Introduction}
An electric current in one conductor can induce a voltage in a neighboring conductor even though the two systems are electrically isolated. This phenomenon - Coulomb drag - has a rich phenomenology and it has been studied extensively in coupled quantum wells since the pioneering experiments by Gramila \textit{et al}.\cite{Gramila1991}  Coulomb drag is a unique transport phenomenon in the sense that the signal is entirely determined by the Coulomb interaction, and thus it provides detailed insight into the many-particle interactions in low-dimensional systems.  Two recent developments have further enhanced the importance of Coulomb drag.  On one hand, samples with graphene layers separated by a nanometer thick boron nitride insulator enter into a new parameter regime, where the interlayer distance is shorter than the mean carrier separation in the two layers \cite{Gorbachev,Titov2013}.  On the other hand, new technologies in sample preparation have allowed the study of drag between one-dimensional (1D) quantum wires, which is 
particularly interesting because of the expected Luttinger liquid formation \cite{Debray2001,Laroche2013,Chen2013}, thus making the plethora of existing theoretical predictions accessible to experimental tests (e.g., Refs. [
\onlinecite{Flensberg1998,Nazarov1998,Ponomarenko2000,Mortensen2001,Mortensen2002,Trauzettel2002}]).

\begin{figure}
\includegraphics[width=1.0\columnwidth]{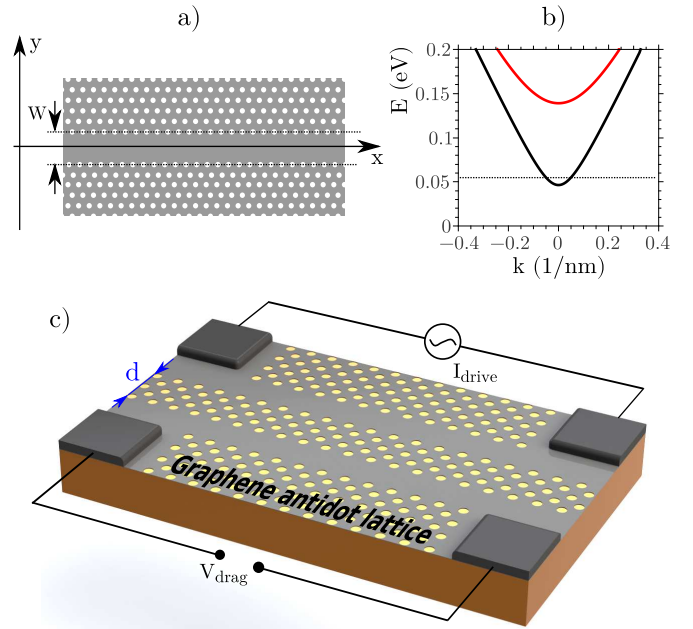}
\caption{(Color online) (a) Schematic illustration of a graphene waveguide (GWG): a region of pristine graphene of width $W$ sandwiched between regions of GALs.
(b) Dispersion relation of GWG with $W = 20$ nm; black dotted line shows a representative Fermi level $E_F = 0.054$ eV, which corresponds to the charge density $n \simeq 3\times 10^{11}\text{cm}^{-2}$.
(c) Coulomb drag setup: two parallel GWGs separated by the region of GAL of the width $d$.}
\label{Fig:setup}
\end{figure}

In the present paper we introduce and analyze a device concept which allows one to study Coulomb drag in one-dimensional graphene systems (see Fig. 1) in a technologically favorable geometry.  The device consists of two graphene waveguides (GWGs), defined with the help of graphene antidot lattices (GALs).  The lateral geometry makes an independent contacting of the two waveguides relatively simple, it avoids complicated gatings \cite{Debray2001}, and no difficult vertical integration is required as in stacked geometries \cite{Laroche2013}. Also the graphene waveguide geometry allows one to avoid complications associated with different electronic properties depending on the orientation of the graphene lattice (zigzag or armchair). The boundary conditions utilized in the Dirac model for GAL defined waveguides do not involve the precise atomistic structure of the edges and thus make a unified description possible. The device design is based on the following considerations.
The antidot lattice creates a band gap \cite{Pedersen2008}, which can theoretically reach several hundreds of meV \cite{Furst2009}, and thus effectively separates the two waveguides.  Theoretical estimates show that three to five rows of antidots provide sufficient electrical isolation\cite{Gunst2011}, implying a minimal separation of a few tens of nanometers.  The waveguides defined via GALs have been shown to have good conduction properties \cite{Pedersen2012}, i.e., they are not so severely affected by disorder as graphene nanoribbons fabricated via an etching process \cite{Chen2013}.  A number of experimental techniques are available for the fabrication of GALs, including block-copolymer \cite{Bai2010} and nanosphere \cite{Wang2013} masks, ion beam etching \cite{Eroms2009} and \textit{e}-beam lithography \cite{Begliarbekov2011}.

The proposed device geometry is, in addition to studies of Coulomb drag, highly relevant to other studies of coupled one-dimensional (1D) structures based on graphene. For example, the propagation of plasmons \cite{Nikitin2012,Christensen2012}, or the effect of a van der Waals interaction \cite{Drosdoff2014} have been investigated recently both theoretically and experimentally in similar systems.

The paper is organized as follows. In Sec. \ref{Sec:Model} we describe three basic ingredients entering our calculations: the model for GAL waveguides, the Coulomb drag theory, and the evaluation of the dielectric function in the random phase approximation (RPA). Section \ref{Sec:Results} presents our numerical results and conclusions, which are summarized in Sec. \ref{Sec:Conclusions}.

\section{Model}
\label{Sec:Model}
\paragraph{Graphene antidot lattice waveguide.}
Low energy excitations in graphene waveguides can be modeled by the Dirac equation with a mass term $m(y)$, which describes the region of graphene sheet with anti-dots \cite{Pedersen2012}, i.e. $m(y)>0, |y|>W/2$, where $W$ is a width of the waveguide [see Fig. \ref{Fig:setup} (a)]. Thus we have to solve the Schr\"odinger equation $\hat{H}\psi(\mathbf{r})=E\psi(\mathbf{r})$, with the Hamiltonian
\begin{equation}
\hat{H} = \hbar v_F
 \left(
  \begin{array}{cc}
   m(y) & -i\frac{\partial}{\partial x} - \frac{\partial}{\partial y}\\\
   -i\frac{\partial}{\partial x} + \frac{\partial}{\partial y} & -m(y)\\
  \end{array}
 \right).
 \label{Hamiltonian}
\end{equation}
Due to the translational symmetry in the $x$ direction, the solution of Eq.(\ref{Hamiltonian}) can be written in the form
\begin{equation}
\psi_n(\mathbf{r})=e^{ikx}
\left(
\begin{array}{c}
\phi_a(y)\\
\phi_b(y)\\
\end{array}
\right).
\end{equation}
We assume that the bang gap produced by GALs is much larger than the Fermi energy, i.e., $E_g^{\text{GAL}} \gg E_F$, which is mathematically expressed as $m \to \infty, |y|>W/2$ (the infinite mass limit). Then, by applying the Berry-Mondragon boundary conditions \cite{Berry1987}, we get the wave-function
\begin{equation}
 \psi_n(\mathbf{r})=\frac{1}{2\sqrt{W}\sqrt{L}}e^{i k x}
 \left(
 \begin{array}{c}
  se^{i\theta} e^{-ik_n y} + e^{ik_n y}\\
  se^{i\theta} e^{ik_n y}  + e^{-ik_n y}
 \end{array}
 \right),\label{Psi_K}
\end{equation}
where $\theta_{k_n, k} = \arctan(k_n/k)$ and $s = \text{sgn}(E)$. The energy dispersion is given by a set of subbands
\begin{equation}
 E_n(k) = s\hbar v_F \sqrt{k^2 + k_n^2}, \quad k_n = \frac{\pi}{W}\left(n+\frac{1}{2}\right).
\end{equation}
The lowest energy excitations can be approximated by quadratic dispersion
\begin{equation}
 E_k \equiv E_0(k) = \hbar v_F \sqrt{k_0^2+k^2} \approx \frac{E_g}{2} + \frac{\hbar^2k^2}{2m^\star}
\end{equation}
with the effective mass $m^\star = k_0 \hbar/v_F$ and the band gap $E_g = 2\hbar v_F k_0$. If the Fermi energy lies in the lowest subband, the density of carriers  is  $n = \frac{g_s g_v}{\pi W}k_{F}$, where $g_s = g_v =2$ is a spin and valley degeneracy and $k_{F}$ is a Fermi wave vector. We emphasize that even though the appropriate dispersion is parabolic, the pseudospin nature of graphene permeates in the calculations due to the wavefunction overlap factor discussed below. 
\begin{figure}
\includegraphics[width=1.0\columnwidth]{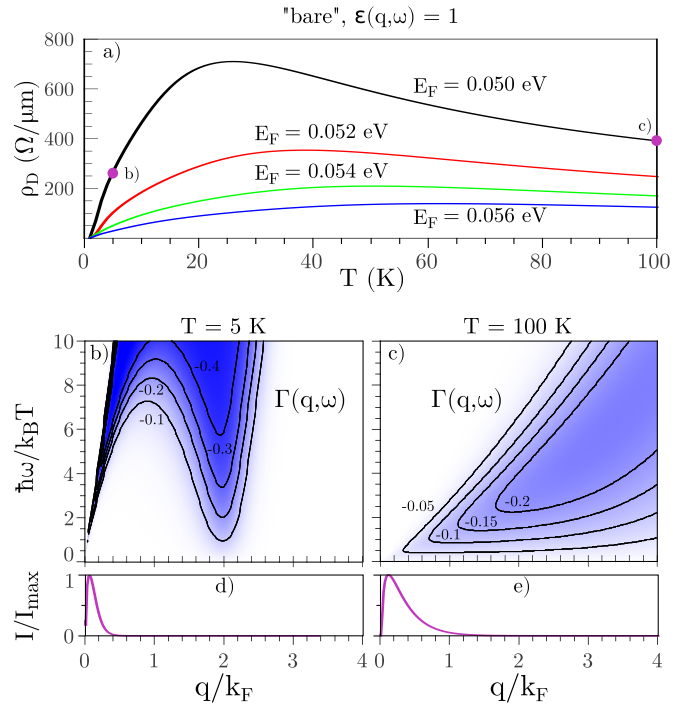}
\caption{(Color online) (a) Drag resistivity between equal ballistic waveguides as a function of temperature for different chemical potentials. The width of the waveguides $W_1 = W_2 = 20$ nm and $d=40$ nm. Nonlinear susceptibility at different (b) $T = 5$ K  and (c) $T = 100$ K temperatures of the system and $E_F = 0.050$ eV. (d), (e) Normalized drag intensity calculated using Eq. (\ref{drag_I}).}
\label{Fig:r21_bare}
\end{figure}
\paragraph{Drag calculation.}
We use the standard expression for the drag resistivity, where the subsystem interaction is taken into account perturbatively up to second order \cite{Jauho1993,Flensberg1995,Kamenev1995},
\begin{align}
 \rho_{21} = \frac{\hbar^2}{16\pi e_1 e_2 n_1 n_2 k_B T}\frac{1}{W}&\int_{-\infty}^\infty \frac{dq}{2\pi}\int_{-\infty}^\infty
 d\omega |U_{12}(q,\omega)|^2 \nonumber\\
 &\times\frac{\varGamma_{1}(q,\omega)\varGamma_{2}(q,\omega)}{\sinh^2(\frac{\hbar\omega}{2k_B T})},
 \label{r21}
\end{align}
where the subscript $i=1,2$ defines the waveguide and $T$ is the temperature, $\varGamma_{i}(q,\omega)$ is the nonlinear susceptibility and $U_{12}(q,\omega)$ is the Fourier component of the screened interwaveguide Coulomb interaction.

\begin{figure}
\includegraphics[width=1.0\columnwidth]{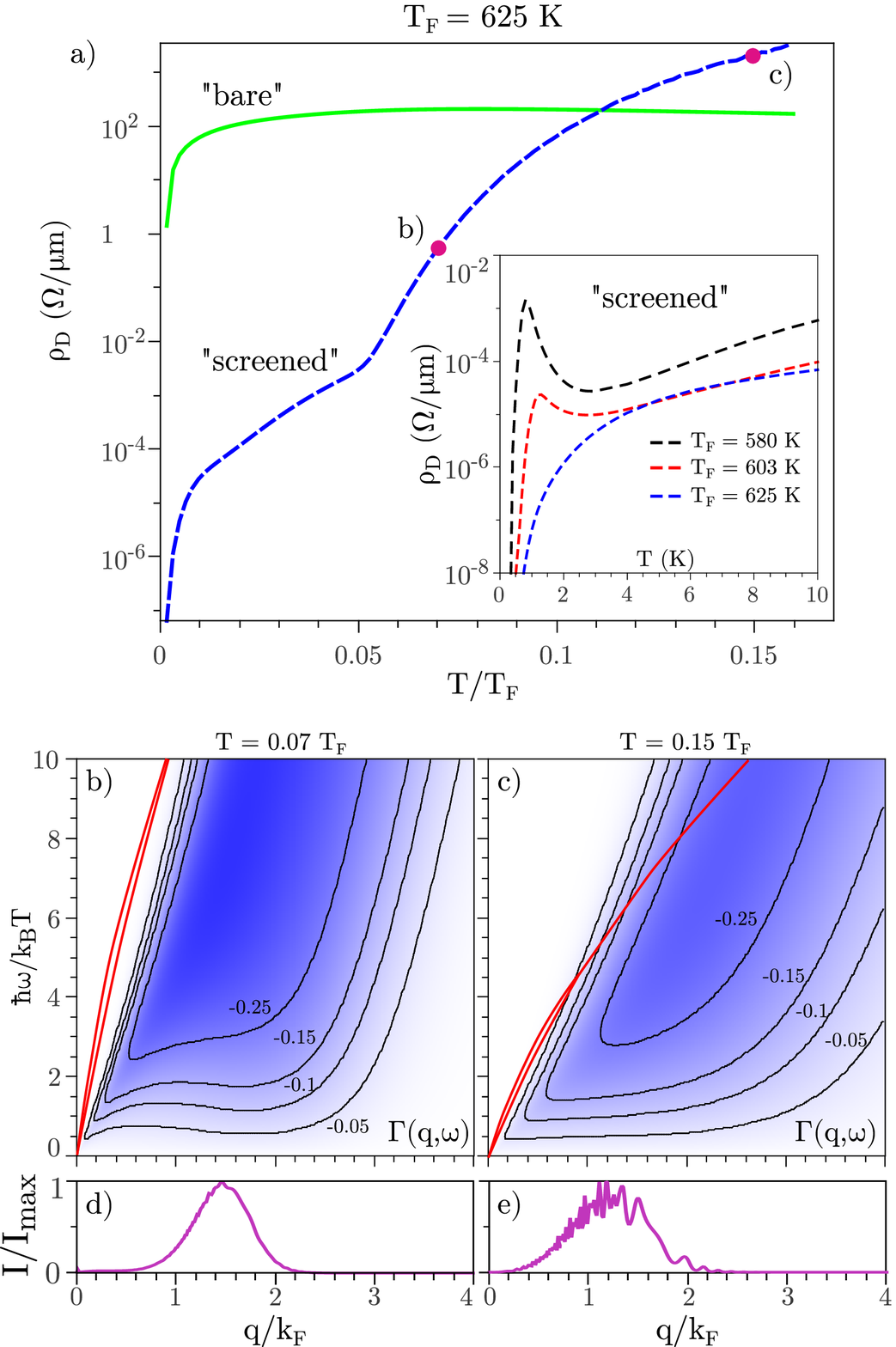}
\caption{(Color online) Drag resistivity as a function of temperature with unscreened (green solid line, the same as in Fig. \ref{Fig:r21_bare}) and screened (blue dashed line) Coulomb interaction. Pink dots on the curve correspond to the temperature points examined on (b) and (c).
Inset: Low-temperature behavior of $\rho_D$. (b), c) The nonlinear susceptibility calculated at different temperatures. Red curves show dispersion $\omega(q)$ of the plasmon modes. (d), (e) The normalized drag intensity.}
\label{Fig:r21_screened}
\end{figure}

In what follows we consider only the lowest subband.
This approximation can be justified by the following arguments.
First, for the parameters chosen for the calculations, namely $E_g \gg k_B T$, the contribution from the interband transition is small compared to the intraband contribution.
Also, an analysis of Eq. (\ref{r21}) shows that the drag resistivity decreases rapidly within increasing Fermi level. If the charge densities are equal in both waveguides,  the drag resistivity scales by the factor $1/n^2$.
Moreover, as we show in detail below, the dominant contribution to $\rho_{21}$ comes from backscattering with momentum transfer $k_F \lesssim q \lesssim 2k_F$ (see Fig. \ref{Fig:r21_screened}). In this case the interaction between waveguides, described by the $U_{12}(q,\omega)$ term, decays rapidly with an increase of $E_F$.
Therefore in order to get a measurable signal, one has to operate at low doping, which corresponds to the Fermi level located in the vicinity of the lowest subband edge.

The nonlinear susceptibility, which describes a response of the charge density to an external potential, is given in the Boltzmann limit (weak disorder) by \cite{Lunde2005}
\begin{align}
 &\varGamma_{i}(q,\omega) = \frac{2\pi e_i g_s g_v}{\hbar\mu_{\text{tr},i}} \int_{-\frac{\pi}{a}}^{\frac{\pi}{a}} \frac{dk}{2\pi} \delta(E_{k}-E_{k+q}-\hbar\omega)  \label{Gamma} \\
  &\times [f(E_{k})-f(E_{k+q})][\tau_{k+q}v_{k+q}-\tau_{k}v_{k}]F(k, k+q), \nonumber
\end{align}
where $\tau_k$ is the transport scattering time,  $\mu_{\text{tr},i}$ is a mobility in a sense that $j=en\mu_{\text{tr}} E$, $v_k = \frac{1}{\hbar}\frac{\partial E(k)}{\partial k}$ is a group velocity, and $f(E)$ is the Fermi-Dirac distribution function. The function $F(k,k+q) = [1+\cos(\theta_{k_0,k+q}-\theta_{k_0,k})]/2$ is the wave-function overlap, which stems from the calculation of the Coulomb interaction matrix element $\langle k,k+q|V(\mathbf{r}_1,\mathbf{r}_2)|k+q,k \rangle$.

In general, the transport scattering time is a function of momentum (or energy) $\tau = \tau(\mathbf{k})$. In the low-temperature limit, $T \ll T_F$, which we consider here, the drag resistivity is not sensitive to the precise functional dependence of $\tau(\mathbf{k})$ \cite{Carrega2012}, so that the relaxation time approximation $\tau(\mathbf{k}) \simeq \tau_F = \text{const}$ can be employed \cite{Narozhny2012,Amorim2012}. Due to the delta function in Eq. (\ref{Gamma}), the integral can be evaluated analytically
\begin{equation}
 \varGamma(q,\omega) = \text{sgn}(q)\frac{k_0 g_s g_v}{\hbar v_F} [f(E_{k_s})-f(E_{k_s+q})]  F(k_s, k_s+q),
\end{equation}
where we used that $\mu_{\text{tr}} = e\tau/m^\star$ and
\begin{equation}
k_s = -\frac{k_0}{q}\frac{\omega}{v_F} - \frac{1}{2}q
\end{equation}
is a root of the equation $E_{k_s}-E_{k_s+q}-\hbar\omega = 0$.
\paragraph{Screening.}
The dynamically screened interwaveguide Coulomb interaction is
\begin{equation}
U_{12}(q,\omega) = \frac{V_{12}(q)}{\epsilon(q,\omega)},
\end{equation}
where $\epsilon(q,\omega)$ is the dielectric function calculated within the random phase approximation \cite{Jauho1993,Flensberg1995,Kamenev1995},
\begin{align}
 \epsilon(q,\omega) &= (1-V_{11}(q)\Pi_{11}(q,\omega))(1-V_{22}(q)\Pi_{22}(q,\omega)) \nonumber \\
 &-V_{21}(q)\Pi_{11}(q,\omega)V_{12}(q)\Pi_{22}(q,\omega)
\end{align}
and $V_{ij}(q)$ are the 1D Fourier components of the bare Coulomb interaction:
\begin{equation}
 V_{ij}(q) = \frac{e_i e_j}{2\pi\epsilon_r\epsilon_0}\int \frac{dy_1}{W_i} \int \frac{dy_2}{W_j} K_0(q\lvert y_1-y_2 \lvert),
\end{equation}
where $K_0(y)$ is a zero-order modified Bessel function, and $\epsilon_r = 2.5$ is the relative dielectric permittivity.
The finite-$T$ polarizability is given by the bare bubble diagram \cite{Brey2007}
\begin{equation}
 \Pi_{mn}(q,\omega) = \frac{g_s g_v}{L} \sum_k \frac{f(E_{k+q}^m)-f(E_{k}^n)}{E_{k+q}^m-E_{k}^n - \hbar(\omega + i\eta)}F(k,k+q),
\end{equation}
where $L$ is a length of the waveguides.

\section{Results}
\label{Sec:Results}
For the sake of simplicity we consider two equal GWGs, i.e., $W_1 = W_2 = 20$ nm, with equal chemical potential and temperature [see Fig. \ref{Fig:setup} (c)]. The distance between GWGs is 40 nm. The band gap inside the waveguide, caused by quantum confinement, is $E_g = 0.092$ eV.
In order to appreciate the role of screening, we first calculate the drag using bare Coulomb interaction, i.e. $\epsilon(q,\omega) = 1$.
Figure \ref{Fig:r21_bare} shows the drag resistivity as a function of temperature for different values of chemical potential. One can see that the value of $\rho_D$ is very sensitive to the value of $E_F$. An increase of $E_F$ by just a few meVs results in a significant drop of the drag resistivity for two reasons. First, a change of the chemical potential induces extra carries in the system, which decreases the drag resistivity because of the factor $n^{-2}$ (for equivalent waveguides) according to Eq. (\ref{r21}). The second reason is related to the fact that scattering with momentum transfer of the
order of $k_F$, described by $U_{21}(q)$, is much smaller for a larger $E_F$.

In the case of an unscreened Coulomb interaction the temperature dependence of the drag resistivity exhibits the following behavior: At small temperatures the drag grows rapidly with increasing $T$ reaching the maximum value at $T \approx 0.05 T_F$. A further increase of the temperature results in either decay of $\rho_D$ (for $E_F = 0.050$ eV) or saturation of its value (for $E_F = 0.056$ eV). The explanation for this behavior is based on a phase-space consideration of $\Gamma(q,\omega)$ function, as we now discuss.

Figures \ref{Fig:r21_bare} (b) and \ref{Fig:r21_bare} (c) show the nonlinear susceptibility $\Gamma(q,\omega)$ as a function of transferred momentum and energy. At low temperatures [$T = 5$ K, Fig. \ref{Fig:r21_bare} (b)) there are two types of excitations available: (i) forward scattering with a small momentum $q \to 0$ and (ii) backscattering with momentum transfer $q \simeq 2k_F$.
[Note that scattering with a momentum around $q=k_F$ requires a large energy transfer and is therefore suppressed due to the factor $\sinh^{-2}(\hbar\omega/2k_B T)$ in Eq. (\ref{r21}).] 
Even though there is much more phase space available around $q = 2k_F$, the forward scattering with a small momentum transfer produces a dominant contribution to the drag, which can be shown by calculating the drag intensity
\begin{equation}
I(q) = \int_{-\infty}^\infty
 d\omega \frac{|U_{12}(q,\omega)|^2 \varGamma_{1}(q,\omega)\varGamma_{2}(q,\omega)}{\sinh^{2}(\hbar\omega/2k_B T)}
 \label{drag_I}
\end{equation}
as depicted in Fig. \ref{Fig:r21_bare} (d).
With an increase of the temperature [$T = 100$ K, Fig.\ref{Fig:r21_bare} (c)] the phase space in between $q=0$ and $q=2k_F$ is distributed almost evenly for low energy excitations. Taking into account that the matrix element of the bare Coulomb interaction grows rapidly with $q \to 0$, the forward scattering is also a dominant process at high temperatures as shown in Fig. \ref{Fig:r21_bare} (e).

Next we analyze the effect of screening on the drag resistivity.
For this purpose we compare the drag resistivity calculated with the bare and the screened Coulomb interaction [see Fig. \ref{Fig:r21_screened} (a)].
The increase of ``screened'' $\rho_D$ with increasing temperature can be understood using the same phase-space arguments as in the case of the drag calculated with bare Coulomb interaction. However at small temperatures $\rho_D^{\text{screened}} \ll \rho_D^{\text{bare}}$.
This is an intuitively expected result, since screening normally lowers the interaction and hence suppresses the drag.
With a further increase of the temperature the relation between the drag calculated with the bare and screened interaction becomes opposite $\rho_D^{\text{screened}} \gg \rho_D^{\text{bare}}$.
We attribute this behavior to a plasmon-mediated enhancement of the Coulomb drag.
Since the drag depends on the screened Coulomb interaction $U_{12}(q,\omega) = V_{12}(q,\omega)/\epsilon(q,\omega)$,
for a certain $\omega(q)$ corresponding to a plasmon mode, the dielectric function tends to zero, $\text{Re}[\epsilon(q,\omega)] \to 0$, which results in a large  $U_{12}(q,\omega)$ and,  in turn,  increases the drag.
Plasmon enhancement of the drag has been considered for coupled quantum wells in Refs. \cite{Flensberg1994,Hu1995,Hill1997} and for two-dimensional (2D) graphene in Ref. \cite{Badalyan2012}.
Thus, we need the plasmon dispersion for the coupled graphene waveguides, i.e., the solutions of $\text{Re}[\epsilon(q,\omega)]=0$. As it is shown in Figs. \ref{Fig:r21_screened} (b) and \ref{Fig:r21_screened} (c) (red solid lines),  two plasmon modes are supported: the out-of-phase (acoustic) $\omega^-(q)$ and in-phase (optic) $\omega^+(q)$ plasmon modes.
At small $q$ the modes are energy resolved and $\omega^+(q)>\omega^-(q)$, while at large $q$ the two branches merge. These coupled plasmon modes are similar to those calculated for the case of two graphene nanoribbons \cite{Christensen2012,Villegas2013}.
As it is shown in Fig. \ref{Fig:r21_screened} (b), at the temperature $T = 0.07 T_F$ the plasmon modes lie outside the particle-hole continuum defined by $\Gamma(q,\omega)$. In this case the screening is effective and therefore $\rho_D^{\text{screened}} \ll \rho_D^{\text{bare}}$.
With an increase of the temperature [$T = 0.15 T_F$, Fig. \ref{Fig:r21_screened} (c)] the nonlinear susceptibility $\Gamma(q,\omega)$ is nonzero at $\omega(q)$'s corresponding to the plasmon modes. 
In this case the Coulomb interaction $U_{12}(q,\omega)$ increases (``antiscreening'') which eventually leads to the enhancement of the drag.
Interestingly screening modifies also the drag intensity. Its maximum lies in between $q=k_F$ and $q=2k_F$ as shown in Figs. \ref{Fig:r21_screened} (d) and \ref{Fig:r21_screened} (e), which means that the backscattering is a dominant process contributing to the drag.

Finally, the inset of Fig. \ref{Fig:r21_screened} (a) shows that the drag resistivity may show an upturn at the very lowest temperatures, depending sensitively on the Fermi energy.  We have not identified a simple physical reason for this behavior: It is a result of a complex interplay between the various factors in the drag formula, Eq.(\ref{r21}). A similar behavior is predicted in drag between Luttinger liquids \cite{Pustilnik2003} and has been recently measured experimentally \cite{Laroche2013}. In Ref.[\onlinecite{Croxall2008}] the upturn of the drag resistivity was also observed in GaAs-AlGaAs electron-hole bilayers. This effect was considered as a signature of exciton superfluidity. Intriguingly, according to their measurements, the upturn may be followed by a downturn of the drag as $T \to 0$. Our calculations exhibit similar trends, but arise here from single particle excitations within the Fermi liquid theory.

\begin{figure}
\includegraphics[width=0.8\columnwidth]{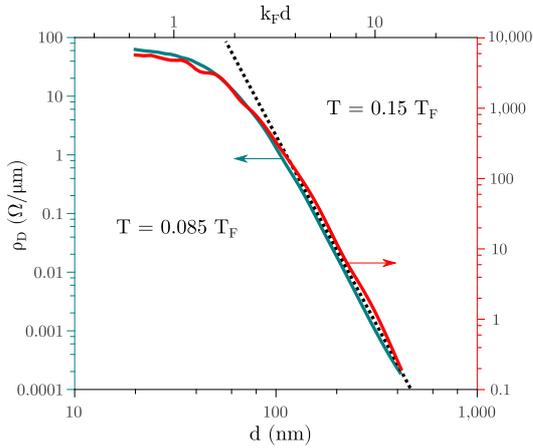}
\caption{(Color online) Drag resistivity between two identical ($W_1 = W_2 = 20$ nm) GWGs as a function of distance $d$ between them calculated at $T = 0.085T_F$ (green line, left axis) and $T = 0.15 T_F$ (red line, right axis).
The Fermi temperature is $T_F = 580$ K and $k_F = 0.0318$ 1/nm. The Black dotted line illustrates asymptotic behavior in the regime $k_F d > 1$.}
\label{Fig:r21(d)}
\end{figure}
Finally, we investigate the interwaveguide distance dependence of the drag, which is depicted in Fig. \ref{Fig:r21(d)}.  These calculations have been carried out at two representative temperatures $T = 0.085T_F$ and $T = 0.15T_F$, which correspond to the screened and enhanced Coulomb interaction respectively. Both curves being properly scaled have approximately the same functional dependence. However, in contrast to the case of 2D graphene sheets, where theoretical predictions \cite{Katsnelson2011,Tse2007} and experimental measurements \cite{Gorbachev} show a $\rho_D \sim d^{-4}$ dependence, we find that in the weak coupling regime $k_F d > 1$ the distance dependence of the drag between two 1D graphene wires follows $\rho_D \sim d^{-n}$, where $n = 6.0 \pm 0.5$.

\section{Summary}
\label{Sec:Conclusions}
In the present paper we have studied the Coulomb drag between graphene waveguides, defined with the help of a graphene antidot lattice. The energy dispersion of GWGs was calculated using the Dirac model with the effective mass term. Using the lowest-subband approximation we compute the drag resistivity. We showed that despite the relatively large interwaveguide separations required for isolated GWGs, the magnitude of Coulomb drag resistivity is in the experimentally measurable range. By performing a detailed analysis of the RPA screening, we found that plasmons provide a significant enhancement of the drag at temperatures $T>0.1T_F$.  At low temperatures the drag resistivity may exhibit a complicated behavior, namely, the upturn of the drag which is always followed by downturn. Finally we showed that in the weak coupling regime the dependence of the drag on interwaveguide separation has $\rho_D \sim d^{-n}$ asymptotic with $n \simeq 6$.  We believe that the device concept suggested here is quite versatile, and may function 
as a platform for many other investigations.

\begin{acknowledgements}
We thank S. Badalyan for insightful remarks.
The Center-of-Excellence CNG is sponsored by the Danish National Research Foundation, Project No. DNRF58.
\end{acknowledgements}

\end{document}